\documentclass[twocolumn,pre,showpacs,superscriptaddress]{revtex4}
\pdfoutput=1
\usepackage{graphicx}%
\usepackage{psfrag}
\usepackage{dcolumn}
\usepackage{amsmath, chemarrow}
\usepackage{amssymb, bm}
\usepackage{latexsym}
\usepackage{hyperref}
\usepackage{color}
\usepackage{wasysym}
\usepackage{mathtools}
\begin{document}

\def\bea{\begin{eqnarray}}
\def\eea{\end{eqnarray}}
\def\beq{\begin{equation}}
\def\eeq{\end{equation}}
\def\f{\frac}
\def\k{\kappa}
\def\e{\epsilon}
\def\ve{\varepsilon}
\def\be{\beta}
\def\D{\Delta}
\def\h{\theta}
\def\t{\tau}
\def\a{\alpha}

\def\ve{\varepsilon}
\def\fv{{\bf{f}}}
\def\fm{\bf{f}_m}
\def\zh{\hat{z}}
\def\yh{\hat{y}}
\def\xh{\hat{x}}
\def\km{k_{m}}

\def\cDa{d\varsigma}
\def\cD{{\cal D}[x]}
\def\cL{{\cal L}}
\def\cLo{{\cal L}_0}
\def\cLa{{\cal L}_1}

\def\Re{{\rm Re}}
\def\sj{\sum_{j=1}^2}
\def\rk{\rho^{ (k) }}
\def\rek{\rho^{ (1) }}
\def\cek{C^{ (1) }}
\def\rz{\rho^{ (0) }}
\def\rt{\rho^{ (2) }}
\def\rtb{\bar \rho^{ (2) }}
\def\trk{\tilde\rho^{ (k) }}
\def\trek{\tilde\rho^{ (1) }}
\def\trz{\tilde\rho^{ (0) }}
\def\trt{\tilde\rho^{ (2) }}
\def\r{\rho}
\def\tD{\tilde {D}}

\def\s{\sigma}
\def\kb{k_B}
\def\la{\langle}
\def\ra{\rangle}
\def\nn{\nonumber}
\def\up{\uparrow}
\def\dn{\downarrow}
\def\S{\Sigma}
\def\dg{\dagger}
\def\d{\delta}
\def\p{\partial}
\def\l{\lambda}
\def\L{\Lambda}
\def\G{\Gamma}
\def\o{\Omega}
\def\w{\omega}
\def\g{\gamma}

\def\noi{\noindent}
\def\a{\alpha}
\def\d{\delta}
\def\p{\partial} 

\def\la{\langle}
\def\ra{\rangle}
\def\e{\epsilon}
\def\n{\eta}
\def\g{\gamma}
\def\rv{{\bf r}}
\def\tv{{\bf t}}
\def\on{{\omega_{\rm on}}}
\def\off{{\omega_{\rm off}}}
\def\break#1{\pagebreak \vspace*{#1}}
\def\hf{\frac{1}{2}}

\def\rcurs{r_{ij}}

\title{Forced desorption of semiflexible polymers,  adsorbed and driven by molecular motors}
\author{Abhishek Chaudhuri}
\email{abhishek@iisermohali.ac.in} 
\affiliation{
Indian Institute of Science Education and Research Mohali, Knowledge City, Sector 81, SAS Nagar - 140306, Punjab, India.
}
\author{Debasish Chaudhuri}
\email{debc@iith.ac.in}
\affiliation{Indian Institute of Technology Hyderabad, 
Yeddumailaram 502205,
Telangana, India}


\date{\today}

\begin{abstract}
We formulate and characterize a model to describe dynamics of semiflexible polymers in the presence of activity due to motor proteins attached irreversibly to a substrate, and a transverse pulling force acting on one end of the filament. The stochastic binding-unbinding of the motor proteins and their ability to move along the polymer,  generates active forces. 
As the pulling force reaches a threshold value, the polymer eventually desorbs from the substrate. Performing molecular dynamics simulations of the polymer in presence of a Langevin heat bath, and stochastic motor activity, we obtain desorption phase diagrams.
The correlation time for fluctuations in desorbed fraction increases as one approaches complete desorption, captured quantitatively by a power law spectral density.
 We present theoretical analysis of the phase diagram using mean field approximations in the weakly bending limit of the polymer and performing linear stability analysis.  This predicts  increase in the desorption force with the polymer bending rigidity, active velocity and processivity of the motor proteins to capture the main features of the simulation results.
\end{abstract}

\pacs{05.20.-y, 36.20.-r, 87.15.-v }

\maketitle

\section{introduction}
Cytoskeleton in the cell comprises of semiflexible protein filaments, cross-linkers and motor-proteins, and is maintained continuously out of equilibrium. Each family of motor proteins, when coupled to their type-specific filamentous tracks, can hydrolyze chemical fuel (ATP), generating motion and stresses in the cell~\cite{Chowdhury2013a,Vale2003,Fletcher2010,Huber2013}. This active meshwork provides the cell its mechanical stability~\cite{Gardel2004,MacKintosh1995}, tracks for intra-cellular locomotion, controlling cell-motility~\cite{Goode2000,Howard2001}, as well as organizing signalling platforms on the cell membrane and endocytosis~\cite{Gowrishankar2012,AChaudhuri2011}.  
Single molecule experiments on motor proteins revealed mechanism of force generation, force-velocity relations, and dependence of motion on ATP concentration~\cite{Oiwa1990, Schnitzer2000a}. Collective action of molecular motors lead to interesting dynamics like bidirectional motion and spontaneous 
oscillations~\cite{Guerin2010,Holzbaur2010,Badoual2002,Vilfan1998}. 
The transport of cargo in one dimension (1D)
by multiple motors has also attracted much attention and the response to external opposing forces have been obtained~\cite{Klumpp2008,Leduc2010,Scharrel2014}. 

A plethora of individual and collective physical properties of cytoskeletal filaments were obtained from the study of {\em in vitro} gliding assays, in which F-actins or microtubules move on a two dimensional substrate decorated by myosin or kinesin motors, both experimentally~\cite{Kron1986,Bourdieu1995,Bourdieu1995d,DeBeer1997a,Schaller2010,Sumino2012} and theoretically~\cite{Yamaoka2012,Kierfeld2008b,Kraikivski2006,Banerjee2011,Vilfan2009,Banerjee2011}. Recent experiments on molecular motor assays revealed formation of spiral defects and loops of actively moving filaments driven by motor proteins~\cite{Bourdieu1995d,Liu2011}. In gliding assays, one end of motor proteins are irreversibly attached to a two dimensional substrate. The other end binds to the filaments and actively forces them to move parallel to the substrate. In this context, it is important to understand the response to external forces of these gliding filaments actively driven by the molecular motors. 

In this paper, we consider a semiflexible filament floating on a molecular motor assay in presence of a pulling force acting on one end of the filament in a direction perpendicular to the substrate. The filament is actively captured and driven parallel to the substrate by molecular motors. 
The situation is akin to {\em in vivo} microtubules, one end of which is captured and actively driven by motor proteins at the cell cortex while forces act on the other end attached to the microtubule originating centre~\cite{Grill2005}. 
A passive counter-part of this problem is peeling of semiflexible polymers from adhesive surfaces~\cite{Oyharcabal2005a,Benetatos2003,Kierfeld2006,Friedsam2004,Cui2003,Aliee2008,Heussinger2007}. Such peeling experiments have been shown to be important in quantifying the strengths of actomyosin rigor bonds, in absence of ATP driven activity, measured by pulling F-actins off myosin coated substrate using optical tweezers~\cite{Nishizaka2000}. Rupture of multiple bonds have also been studied in the context of cell adhesion~\cite{Seifert1990,Seifert2000,Erdmann2004,Erdmann2004a,Smith2007} and the unzipping of DNA~\cite{Marenduzzo2001,Kapri2004,Kumar2013}. Semiflexible polymers themselves are known to show interesting mechanical and dynamic properties~\cite{Dhar2002, Ranjith2002, Ranjith2005, Hallatschek2005, Chaudhuri2007a, Ghosh2009, Liverpool1998}. 

We use Langevin dynamics simulation and theortical mean field analysis to quantify the dynamics, and hence the response, of the semiflexible filament and the molecular motors, under the transverse pulling force. For simplicity, we assume that the motor proteins, which undergo attachment-detachment kinetics, are arranged uniformly on a one-dimensional substrate. With increase in the transverse pulling force applied to one end of the polymer, it undergoes a non-equilibrium {\em continuous} transition at a threshold force, from an adsorbed state to a completely desorbed state. A theorteical mean field analysis of the problem predicts an increase in the threshold force of desorption with increasing bending rigidity of polymer, as well as activity and processivity of molecular motors. The predictions show good agreement with simulation results. We obtain results for biologically relevant parameter values, making our predictions amenable to direct verification in gliding assay experiments.

\section{Model} 
\label{model}
We model the cytoskeletal filaments  as  stretchable semiflexible polymers described by space curve  $\rv(s)$, and local tangent vector $\tv(s) = \p \rv/\p s$ with $s$ denoting a position along the contour of the chain.
The Hamiltonian for a filament of length $L$ is given by \cite{Kierfeld2004} 
\bea
H =  \f{1}{2}\int_0^L ds \left[ \k \left(\f{\p^2 \rv}{\p s^2}\right)^2 +  A  \left(\f{\p \rv}{\p s} \right)^2 \right] 
\label{hamiltonian}
\eea
where  $\k$ is the bending rigidity, $A$ is the bond strength.  This model reduces to the worm like chain model in the limit of unstretchable bonds with $[\tv(s)]^2 =1$.

The motor proteins are modeled as elastic linkers. The {\em tail} end of $i$-th motor protein is attached irreversibly to the substrate at position $\rv_0^i=(x_0^i,0)$. The {\em head} end is free to attach (detach) to (from) the filament. It attaches to a segment of the filament if it lies within a capture radius $r_c$ with an attachment rate $\on$. When attached, the motor head either moves along the filament or gets detached from it with a rate $\off$. The $i$-th molecular motor when attached to the polymer at a position $\rv(s)$ exerts an elastic force 
$\fv_m = -\km (\rv(s) - \rv_0^i) = -k_m\delta{\bf r}$, attracting the polymer segment towards itself. 
The amplitude of this force is the load $f_l=|\fv_m|$ on the head of a molecular motor.
In their active state, the attached end of motor proteins move along the filament towards one of its ends (plus end for kinesins walking on microtubules) with active relative velocity~[see Fig.\ref{fzfs}($a$)] 
\bea
v^a_t(f_t) = \f{-v_0}{1+d_0\exp(f_t/f_s)}
\label{velc} 
\eea
where $f_t=-\fv_m.\tv$, $d_0 = 0.01$ and $f_s = 1.16$ pN is the stall force (parameters corresponding to kinesin molecule, see Appendix~\ref{appendixB}). The negative sign is chosen
so that the motors move towards the $s=0$ end of the polymer.
This ATP-driven motion of the motor heads, via their elastic nature generate an active force on the filament in a direction opposite to this movement, resulting in  a sliding motion of the filament with respect to the substrate. The external force $F_z$, applied at the $s=0$ end of the polymer opposes this active force, and at sufficient strength desorbs the polymer from the substrate.  
 

Note that, $\on$ and $\off$ are dependent on the separation of a given polymer segment from the substrate. In our simplified quasi one-dimensional model, the motors are attached uniformly with the density $\rho$ along the $x$-axis and the transverse fluctuations of the polymer and the motor heads are along the $z-$axis (Fig.~\ref{fzfs}(a)). Assuming a Kramer's process we have,
$\off=\w_0 \exp(f_l/f_d)$, where $\w_0$ is the bare off rate, $f_l$ is the load force originating from the elastic extension of the motor spring and $f_d$ is the typical force required to detach the motor head from the polymer segment.
An external force $F_z$, applied at the $s=0$ end of the polymer opposes this active force, and at sufficient strength desorbs the polymer from the substrate.  


\section{Simulation}
To study the full dynamics of the semiflexible polymer under the influence of motor proteins attached to a substrate, and pulled out of the substrate by an external force, we perform molecular dynamics (MD) simulations of the polymer in presence of a Langevin heat bath, and stochastic attachment detachment kinematics of molecular motors.
In simulations, we discretize the semiflexible polymer into a bead-spring chain of $N$ bonds of equilibrium length $\s$, spring constant $A$, 
and finite bending rigidity $\k$ such that the Hamiltonian is
$H = \sum_{n=1}^{N-1} (\k/2 \s) [\tv(n+1) - \tv(n)]^2 + \sum_{n=1}^N (A/2) [b(n)-\s]^2$. 
Here we denoted position of the $n$-th bead by $\rv(n)$, such that the local tangent $\tv(n) = [\rv(n+1)-\rv(n)]/b(n)$ where $b(n)=|\rv(n+1)-\rv(n)|$ is the instantaneous bond length.
In the limit of large $A$,  instantaneous bond lengths $b(n) \approx \s$, and the chain in equilibrium behaves like a worm like chain. 
In addition, we incorporate self avoidance via a Weeks-Chandler-Anderson (WCA) purely repulsive potential between non-bonded polymer beads 
$\be V_{WCA}(\rcurs) = 4[(\s/\rcurs)^{12} - (\s/\rcurs)^6 + 1/4]$ if $\rcurs < 2^{1/6}\s$ and $0$ otherwise, with $\be = 1/\kb T$ the inverse temperature. 

\begin{figure*}[t] 
\begin{center}
\includegraphics[width=\linewidth]{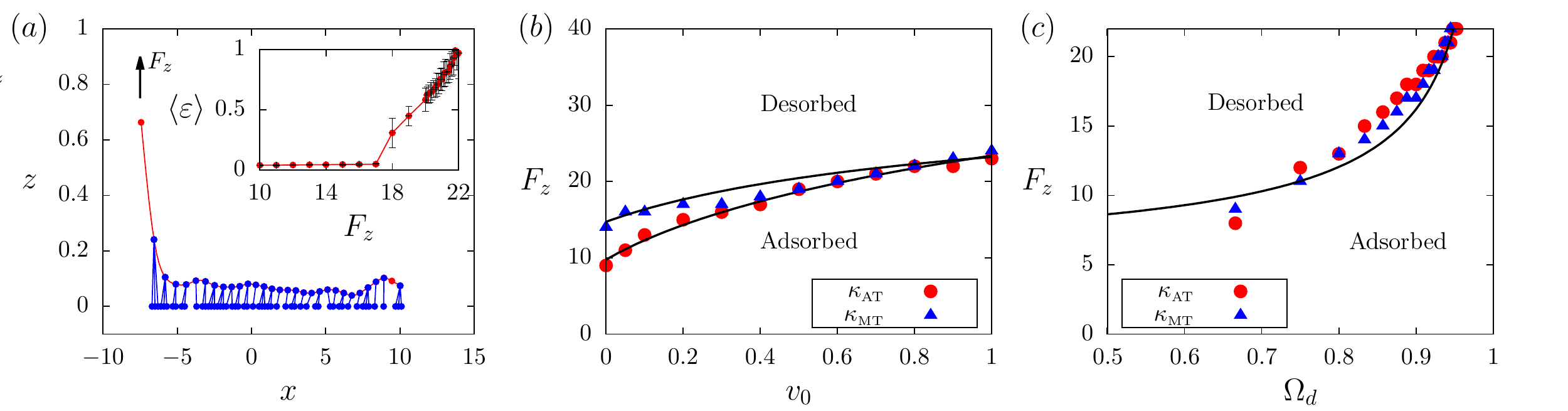} 
\caption{(Color online) Polymer configuration, and adsorption-desorption transitions. 
Units used are: positions in $\mu$m, forces in pN, velocities in $\mu$m/s. 
($a$)~A configuration of semiflexible bead-spring polymer (red beads and lines) actively captured by motor proteins on a substrate. Each motor protein is denoted by a blue line connecting two end-points, one of which is bound irreversibly to the substrate while the other end can attach (detach) to (from) the polymer beads with fixed rates. The end of motor proteins attached to a polymer bead walks towards the trailing end of the polymer on which an external force $F_z$ is applied. This in turn pulls the filament in the opposite direction. 
(Inset) The time averaged desorbed fraction of polymer length $\la \ve \ra$ as a function of $F_z$, where $\ve= \ell/L$ with $\ell$ denoting desorbed length of the polymer. $\la \ve \ra$ remains small up to  $17\,$pN, subsequently showing gradual increase finally desorbing completely at a threshold force $F_z^c = 22$pN. 
($b$)~Phase diagrams depicting adsorption-desorption transition of a stiff (microtubule) and a relatively flexible (F-actin) filament, as a function of
active velocity $v_0$ of motor proteins. Both of them show increase in desorption force with increasing $v_0$, finally merging into a single curve at high enough values of $v_0$. 
Lines are fit to Eq.~\ref{Fztot} with $b^{\prime}_0 = 0.56, b_1^{\prime} = 1/\mu m, b^{\prime}_2 = 5.11$ pN-s/$\mu$m$^3$ and $b^{\prime}_3 = 18.53$ pN/$\mu$m$^2$ for F-actin and $b^{\prime}_0 = 0.384, b_1^{\prime} = 1/\mu m, b^{\prime}_2 = 15.126$ pN-s/$\mu$m$^3$ and $b^{\prime}_3 = 0.755$ pN/$\mu$m$^2$ for microtubule.
$(c)$~Phase diagram for $F_z$ vs $\Omega_d$ for $v_0 = 0.807 \mu$m/s for the same two filaments as in ($b$). 
The line is a fit to Eq.~\ref{FzOd} with the same $b'_0$ as obtained in ($b$) for microtubule.
}
\label{fzfs}
\end{center}
\end{figure*}

\noindent
\begin{table}[t]
\begin{tabular}{l c c}
\hline
Parameters & Definition & Values \\
\hline \\
$\g_w$ & Viscosity of water & $0.001$ pN-s/$\mu$m$^2$\\
$T$ & Temperature & $4.2$ pN-nm/$k_B$ \\
$\alpha$ & Viscous drag & $1$ pN-s/$\mu$m \\
$\s$ & Bond length & $0.5$ $\mu$m \\
$A$ & Spring constant(filament) & $100$ pN/$\mu$m \\
$k_m$ & Spring constant(motor) & $100$ pN/$\mu$m \\
$\r$ & Linear density (motor) & $2.5/\mu$m \\
$\kappa_{\textrm AT}$ & Bending rigidity (Actin) & $0.07$ pN-$\mu$m$^2$ \\ 
$\kappa_{\textrm MT}$ & Bending rigidity ($\mu$tubule) & $21.84$ pN-$\mu$m$^2$ \\ 
$v_0$ & Free motor velocity & $0.8$ $\mu$m/s (K)\\
$\omega_{\textrm on}$ & Attachment rate & $20$/s (K) \\
$\omega_0$ & Bare detachment rate & $1$/s (K) \\
$f_d$ & Detachment force & $6$ pN (K)\\
$f_s$ & Stall force & $1.16$ pN (K) \\ \\
\hline
\end{tabular}
\caption{Various parameters and their typical values used in the simulation. (K) denotes kinesin.}
\label{table-1}
\end{table}

The MD simulations are performed using a velocity-Verlet algorithm in presence of a Langevin heat bath that fixes the temperature at the room temperature value $\kb T=4.2$ pN-nm through a Gaussian white noise $\la \eta_i(t) \eta_j (t') \ra = 2 \a \kb T \d_{ij} \d(t-t')$ with $\a$ denoting viscosity of the environment.  Since the typical environment within a cell is at least one order of magnitude more viscous than water, we choose the viscosity of the medium $\g = 100 \g_w = 0.1$ pN\,s/$\mu$m$^2$, where $\g_w = 0.001$ pN\,s/$\mu$m$^2$ is the viscosity of water. Therefore, the viscous drag $\alpha = 6\pi\g a \approx 1$ pN\,s/$\mu$m on a bond of length $\s=0.5\,\mu$m. The spring constant of the bead-spring system is taken to be fairly large $A = 100$ pN/$\mu$m so that the bond fluctuations are small enough to reproduce known equilibrium statistics~\cite{Dhar2002,Chaudhuri2007a}. 
The persistence length $\l$ of cytoskeletal filaments varies by three orders of magnitude, with $\l=16.7\,\mu$m  for actin filaments, to $\l=5.2$ mm for microtubules~\cite{Gittes1993}.
These correspond to variation of bending rigidity $\k$ from the value $\k_{\rm AT} = 0.07$ pN-$\mu$m$^2$ for F-actins to $\k_{\rm MT} = 21.84$ pN-$\mu$m$^2$ for microtubules. Unless stated otherwise, in our simulations we consider parameters typical for kinesin motors with attachment rate, $\on = 20/$s and a bare detachment rate $\w_0 = 1/$s.  The motors are placed on a 1d line along the $x$-axis with constant coverage density. The spring constant, $k_m$, for kinesin motors lie between $10$-$1000$ pN/$\mu$m. In our simulations, we use $k_m = A=100\, {\rm pN}\, \mu{\rm m}^{-1}$. The detachment force, $f_d = 6$ pN characterizes the force induced enhancement of detachment rates as $\off = \w_0 \exp(f_l/f_d)$ where $f_l$ is the instantaneous load on the molecular motor.  In our simulations, we used the polymer bond-length $\s$ as unit of length, and the typical forces associated with the motor proteins $1\,$pN as the unit of force. 

The unit of time is set by $t_u = \a/A$, and we choose the integration time step $\d t = 0.01 t_u$. 
Attachment- detachment kinematics of motor proteins are performed stochastically with probabilities $\on \d t$ and $\off \d t$ at every time-step. Note that the attachment event is tried only if a filament segment is within the capture radius $r_c\sim\s$, from the {\em equilibrium} position of molecular motors.  Once detached the molecular motors are assumed to relax back to equilibrium configurations immediately. When attached to the polymer, molecular motors move along the polymer following Eq.(\ref{velc}).
In the absence of load, the attached motors move with velocity $v_0$ in the negative $x$ direction along the filament, forcing the polymer to translate towards the positive $x$ direction.

To study the effect of transverse pulling force, $F_z$ is applied in the $z$ direction perpendicular to the substrate and at the trailing end of the polymer (see Fig.~\ref{fzfs} ($a$)). 
We study the influence of the pulling force $F_z$ as we vary (i)~active self-propulsion $v_0$, (ii)~the duty ratio $\o_d$, 
and (iii)~the bending rigidity $\kappa$. 
All the parameters used in the simulations are summarized in Table~\ref{table-1}. 



\begin{figure}[t] 
\begin{center}
\includegraphics[width=0.9\linewidth]{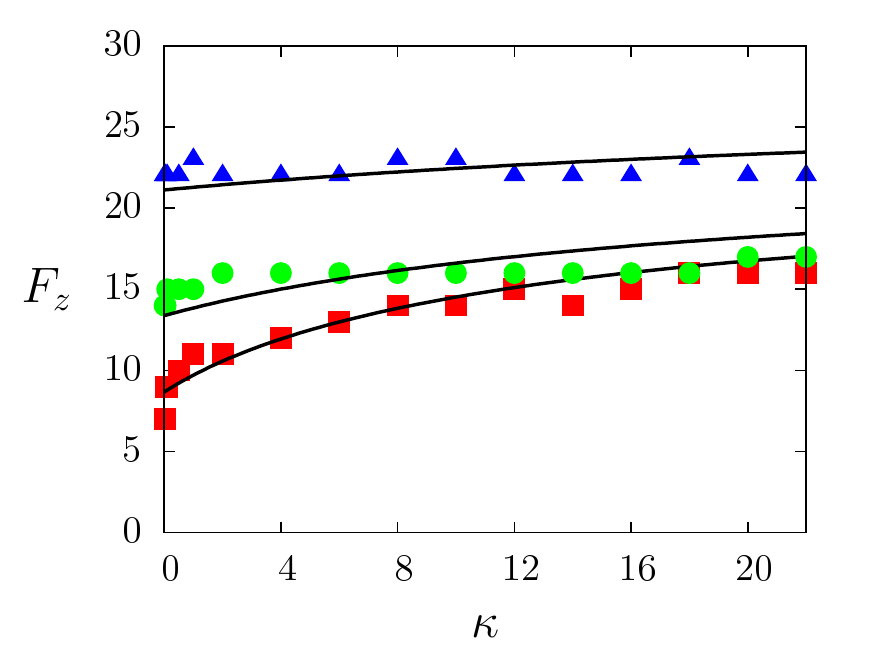} 
\caption{(Color online) 
Phase diagram for $F_z$ (in pN) as a function of $\kappa$ (in pN-$\mu {\rm m}^2$) for a passive system ($v_0 = 0$, $\blacksquare$), intermediate activity ($v_0 = 0.2 \mu$m/s, $\CIRCLE$) and high activity ($v_0 = 0.807 \mu$m/s, $\blacktriangle$).      
Lines are fit to Eq.~\ref{Fztot} with $b^{\prime}_0 = 0.267$, $b_1^{\prime} = 1/\mu m$, $b^{\prime}_3 = 20.97$ pN/$\mu$m$^2$ for $v_0 = 0$, $b^{\prime}_0 = 0.278$, $b_1^{\prime} = 1/\mu m$, $b_2^{\prime} = 28.737$ pN-s/$\mu$m$^3$, $b^{\prime}_3 = 20.97$ pN/$\mu$m$^2$ for $v_0 = 0.2$ and $b^{\prime}_0 = 0.33$, $b_1^{\prime} = 1/\mu m$, $b_2^{\prime} = 28.737$ pN-s/$\mu$m$^3$, $b^{\prime}_3 = 20.26$ pN/$\mu$m$^2$ for $v_0 = 0.807$.
}
\label{fzkappa}
\end{center}
\end{figure}

\section{Results}

In the absence of a transverse pulling force and for processive motors with large duty ratio $\Omega_d=0.95$  ($\on:\w_0=20:1$), desorption of the polymer by stochastic fluctuations is prevented. The polymer stays on the motor protein bed and slides towards positive $x$-axis with a velocity close to $v_0$ characteristic of individual motor proteins. The response of the system to a force applied {\em parallel} to the substrate is characterized in detail in Appendix~\ref{appendixA}. In this specific case, filament bending does not play any role and the filament is well approximated as a one dimensional rigid rod. 


For a transverse pulling force applied on the trailing end, opposing the sliding motion of the filament, we study the transition of the polymer from an adsorbed to a completely desorbed state.
Fig.\ref{fzfs}($a$) shows a typical configuration  of a filament 
having $\k=\k_{\rm MT}$ the bending rigidity of microtubules, on a bed of motor proteins having active velocity $v_0=0.807\, \mu{\rm m}\, s^{-1}$ corresponding to 
kinesins, under external force $F_z$. For further details on modeling kinesin activity see Appendix~\ref{appendixB}. 
The desorption is characterized by a continuous increase in the fraction of desorbed length beyond a threshold force, as shown in the inset of Fig.~\ref{fzfs}(a), or by following the fraction of motor proteins attached to the filament. Note that beyond $F_z = 17$pN in  Fig.~\ref{fzfs}(a), the polymer starts to partially desorb, while the complete desorption occurs at a relatively higher pulling force of $F_z^c=22$pN. The transition from adsorbed to desorbed state occurs {\em smoothly}, like a continuous phase transition. We come back to this point again at the end of this section. We obtain $F_z^c$ with changing $v_0$, $\o_d$ and $\k$, in each case keeping the other parameter values unchanged. This gives us three phase diagrams for adsorption-desorption transition.

  \begin{figure}[t] 
\begin{center}
\includegraphics[width=0.9\linewidth]{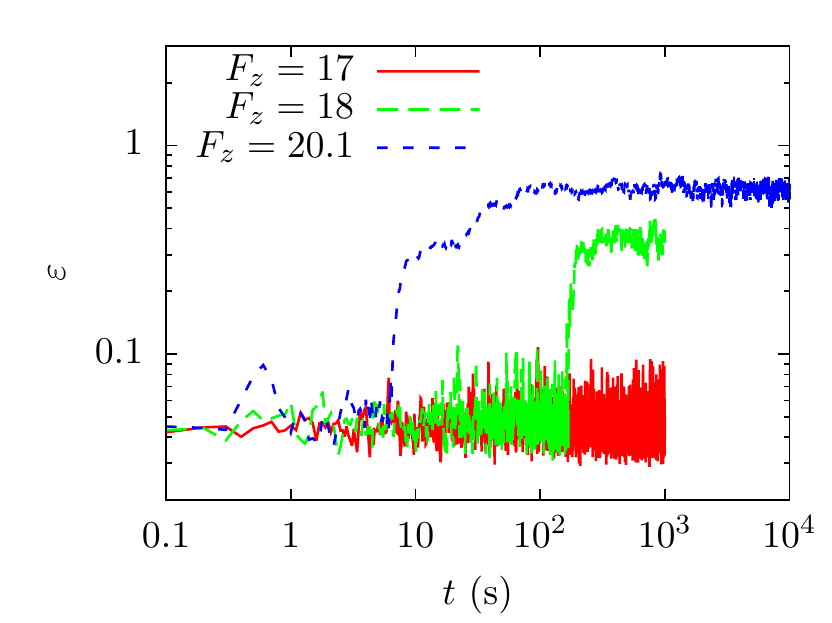} 
\caption{(Color online) 
Time series of the desorbed fraction of microtubule from kinesin bed, at different values of pulling force $F_z$ expressed in units of pN. A longer time series is presented  
at $F_z=20.1$pN, a force value close to the desorption transition.}
\label{tseries}
\end{center}
\end{figure}

Fig.~\ref{fzfs}($b$) shows the dependence of $F_z^c$ on $v_0$, for two different values of bending rigidity of polymer $\k$:  $\k = \k_{\rm AT}$ corresponds to relatively flexible F-actins, and $\k = \k_{\rm MT}$ corresponds to the very rigid limit of microtubules. The simulation results show that both for F-actins and microtubules  the threshold desorption force $F_z^c$  increases  with $v_0$,  to eventually merge together and saturate at large values of $v_0$ in accordance with Eq.~\ref{Fztot}, derived in the following section.  Both the data sets in  Fig.~\ref{fzfs}($b$) fit well with Eq.~\ref{Fztot}.  Fitting parameters are mentioned in the figure captions. 
$F_z^c$ increases with $v_0$ to eventually saturate at large $v_0$ as $F_z^c \sim [1-{\cal C}_1/ ({\cal C}_2 + v_0) ]^{1/2}$, where ${\cal C}_1$,  ${\cal C}_2$ are constants. 
At large enough values of $v_0$ and $\k$, the desorption force is expected to become independent of both~[see Eq.\ref{FzOd}], 
leading to the same $F_z^c$ for F-actins and microtubules for large active velocity $v_0$ of the molecular motors. Thus the two phase boundaries for F-actins and microtubules merge together as $v_0$ approaches $1\mu$m/s.
Fig.~\ref{fzfs}($c$) presents the simulated phase diagram in $F_z^c - \o_d$ plane, calculated with large active motion 
$v_0=0.807\,\mu$m/s. It shows good agreement with mean field results (Eq.~\ref{FzOd}). 

In Fig.~\ref{fzkappa}, we present the dependence of desorption force $F_z^c$ on bending stiffness $\kappa$ of the polymer, for different values of active velocity $v_0$. 
All the data sets fit well to Eq.~\ref{Fztot}.
A {\em passive} semi-flexible polymer ($v_0 = 0$) shows increase in $F_z^c$ with increasing $\kappa$ as $F_z^c \sim [1-{\cal C}'_1/({\cal C}'_2 + \k) ]^{1/2}$, with ${\cal C}'_1$ and ${\cal C}'_2$  are constants. 
This scenario is equivalent to equilibrium desorption of semiflexible filaments from adhesive substrates~\cite{Benetatos2003,Kierfeld2006}.
However, at large values of active velocity of motor-proteins ($v_0 = 0.807\, \mu$m/s), $v_0$ would dominate over $\k$,  giving rise to an essentially $\k$-independent desorption force. 

\begin{figure}[t] 
\begin{center}
\includegraphics[width=0.9\linewidth]{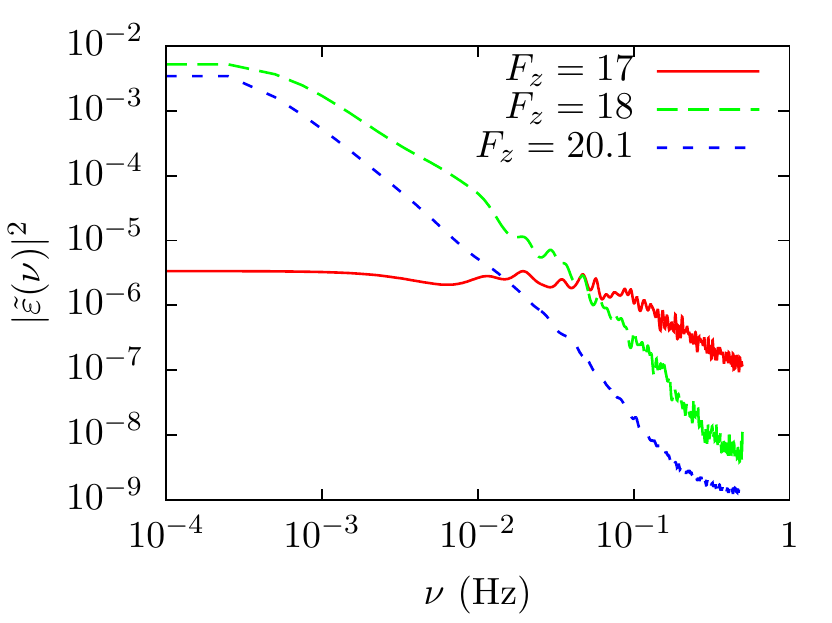} 
\caption{(Color online) 
Power spectrum of the desorbed fraction $|\tilde \varepsilon (\nu)|^2$ obtained from the respective time series as shown in Fig.~\ref{tseries}. The data show power law profiles $1/\nu^\a$ with $\a\approx 1$ at $F_z=17$pN, and $\a\approx 2$ at $F_z=20.1$pN, and reflect an increase in the correlation time as one approaches the desorption transition force $F_z = 22$pN. 
}
\label{spectrum}
\end{center}
\end{figure}

Finally we take a closer look at the desorption process itself. For this purpose we use the time evolution of the order parameter, the desorbed fraction of the polymer, $\varepsilon$. The time series of this stochastic quantity  $\varepsilon(t)$ shows different behavior depending on the value of the applied desorbing force $F_z$ (Fig.\ref{tseries}).  At long time $\varepsilon(t)$ reaches a steady state, with a time-independent mean value. The gradual approach to final steady states with $\varepsilon \approx 0.4$ for $F_z = 18$pN to $\varepsilon \approx 0.6$ for $F_z = 20.1$pN shows how the polymer desorption progresses in time.  Note that the continuous change in average order parameter $\langle \varepsilon \rangle$ with $F_z$ as shown in Fig.~\ref{fzfs}(a) inset, indicates a {\em continuous} non-equilibrium  transition.  The power spectrums presented in Fig.~\ref{spectrum} quantifies the dynamics. This clearly shows that the correlation time of the dynamics increases as one increases $F_z$ towards desorption transition value. The power spectral density shows a $1/\nu^\a$ behavior with $\a \approx 1$ for $F_z \leq 17$pN, and $\a \approx 2$ as one approaches desorption, e.g., see the behavior  at $F_z = 20.1$pN. This is reminiscent of pink noise observed in diverse types of non-equilibrium systems~\cite{Bak1987,Bursac2005}. 


\section{Theory}
The full dynamics is fairly complicated and not tractable analytically. However, simplifications of the original problem is useful to develop analytical insight. In the following section, 
we use small bending approximation for the filament, and mean field approximation for dynamics and external force sharing. Further, the net impact of the molecular motors on the filament is treated in terms of  a local elastic binding to the substrate, and a sliding velocity. Since the transverse force is applied at the end of the polymer opposite to the direction of sliding, a component of the external force acts against the direction of sliding.  In the simulations, these opposing forces result partially into elastic energy stored in the extension of stretchable but stiff bonds, and into building up a tension along the chain. In the following, we assume an unstretchable chain in which the tension alone bears the impact of these two opposing forces.

\subsection{Mean field analysis}
\label{theory} 
In the mean field limit, let us assume that $n_{b,u}$ denote the local density of bound and unbound motors, such that $n_b+n_u=\r$ is a constant. 
In the detached state motors play no role in the dynamics of filament. However, the detached motor heads after performing diffusive relaxation reattaches to a polymer segment within capture radius with a given rate.  
We assume that the attachment of a motor head to the polymer is not instantaneous but happens over a time $t_a$. Further, to attach to the filament, the motor head diffuses over a distance $z$, the local transverse distance of the polymer segment, with a time scale $z^2/D_u$. Therefore, the total time required for the process is $(t_a + z^2/D_u)$, giving rise to an effective attachment rate, $\on = D_u/(t_a D_u + z^2)$. 
Thus the dynamics of unbound motors may be incorporated in terms of the effective $\on$. 
The bound motors generate elastic force per unit length $-\km z\, n_b$ on the filament, and slides the filament with a velocity $-v_t^a(f_t)$ where $f_t \propto F_z/n_b L$.  

The bound motors detach from the filament with rate $\off=\w_0 \exp(f_l/f_d)$, where
the local load force, $f_l = -k_m|\delta{\bf r}|$. This stochastic load originates from three mechanisms: 
(i)\,stochastic binding/unbinding of other motor proteins changing the number of bound motors sharing the load, 
(ii)\,motion of bound head of motor proteins along the filament, 
and finally (iii)\,the external force, $F_z$, acting at the filament end. In the absence of $F_z$, stochastic binding/unbinding  will result in an 
average time-independent 
separation $|\delta{\bf r}|$ and the sliding motion of the filament will hand over binding from one motor to its neighbor without impacting the polymer dynamics on an average. 
Therefore, the average load force would really be due to the external force $F_z$. 
Within mean-field approximation, we assume that $F_z$ acting on the polymer is distributed equally among all bound motors. 
Therefore, one may use $f_l = F_z/n_b L$ to obtain $\off = \w_0 \exp(f/n_b f_d)$, where $f=F_z/L$. 
This remains a good approximation within the weakly bending limit. 

In the limit of small transverse displacements $z(x,t)$ of the filament from $x$-axis, the Hamiltonian (Eq. ~\ref{hamiltonian}) may be approximated as
\bea
{\cal H} = \f{1}{2} \int_0^L ds \left[ \k \left(\f{\p^2 z}{\p x^2}\right)^2 + \t(x,t) \left(\f{\p z}{\p x}\right)^2 \right],
\eea
where instead of a large spring constant $A$ used in the simulations, we use a   local instantaneous tension $\t(x,t)$ in the theory, with $\t(x,t)$ constraining the local bond lengths to a constant value $\s$.
The over-damped motion due to this Hamiltonian is described by $\a_\perp \p_t z = -\d {\cal H}/\d z + \eta(t)$, where $\eta(t)$ is a Gaussian white noise, and $\a_\perp$ viscous friction.
Averaging over the stochastic noise, and incorporating the force due to bound motors, the evolution of the transverse displacement and the attachment-detachment dynamics of the unbound motors is given by,
\bea
\a_\perp\f{\p z}{\p t} &=& -\k \f{\p^4 z}{\p x^4} + \t(x,t) \f{\p^2 z}{\p x^2}
- \km z\, n_b \crcr 
\f{\p n_u}{\p t} &=& \off \r - (\on+\off) n_u.
\label{dyn-1}
\eea
The tension $\t(x,t)$ needs to be determined using the inextensibility constraint. In the weakly bending limit, spatial variation in $\t$ can be neglected~\cite{Hallatschek2005}, considerably simplifying the analysis.
For values of the external force, $F_z$, less than the critical desorption force, the polymer reaches a steady state configuration $z(x)$ (independent of $t$) where it is partially adsorbed.

\subsection{Linear Response} 
To get an estimate of the critical force required to desorb the filament from the substrate, we perform a linear stability analysis by assuming 
a steady state configuration $[z(x),n_b(x)]$ obtained under a fixed external force $F_z$. Let us consider small variations $[\d z(x),\d n_b(x)]$ about it. 
As stated earlier, the attachment rate $\on =D/(t_aD+z^2)$, and detachment rate $\off = \w_0 \exp(f/n_b f_d)$.
Small variations around steady state give $\d \off =  - (\off f/n_b^2 f_d) \d n_b$, $\d \on = - [2 \on z/(t_aD+z^2)] \d z$. 
In practice, a motor protein may attach to a segment of filament only if it lies within a capture radius $r_c$. 
We assume that the segment of filament to which motors may attach remains essentially parallel to the substrate and within a separation $\s$. Thus replacing $r_c$ by $\s$,
$\km z\, \d n_u$ by $\km \s\, \d n_u$, and rewriting $\d \on = - b_1\on \d z$ with $b_1 = 2 \s/(t_aD + \s^2)$, and $\d n_b$ by $-\d n_u$ we obtain from Eq.(\ref{dyn-1}),
\bea
\a_\perp \p_t \d z = (-\k \p_x^4 + \t \p_x^2 - k_m n_b)\d z + \km \s\, \d n_u \crcr 
\p_t \d n_u = \left[\f{\off f}{n_b f_d} - (\on + \off) \right] \d n_u + b_1 \on n_u \d z. \nn 
\label{eq:lindyn}
\eea

As argued before, the total tensile force $\t$ may be expressed in terms of 
active processive motion of the motor proteins as 
$\t 
 = b_2 v^a_t + b_3$, with 
 $b_2$ a undetermined constant. Here the constant $b_3$ denotes the tension due to joint action of  external force $F_z$ and adhesion to substrate by the motor proteins.
 The linear perturbations considered above are variations around a {\em steady state} where we assumed that all tension propagation~\cite{Hallatschek2005} is settled down. 
 Thus a linear dependence of $\t$ on a steady state velocity $v^a_t(F_z)$ is reasonable.

If one considers a small segment of a long polymer, far away from the boundary on which pulling force $F_z$ is applied, boundary conditions would not affect the local
behavior. In this limit, performing a Fourier transform, the evolution of specific modes follows 
$\p_t ( \d z^q, \d n_u^q) = {\cal A}. ( \d z^q, \d n_u^q)$, where the elements of the $2\times 2$ matrix ${\cal A}$ are given by 
${\cal A}_{11} = -[\k q^4 + \t q^2+k_m n_b]$, 
${\cal A}_{12} = k_m \s$,
${\cal A}_{21} = b_1\on n_u $,
and
${\cal A}_{22} = \left[\f{\off f}{n_b f_d} - (\on + \off) \right]$.

In the large $q$ limit, the unstable modes of the above linearized dynamics identifies the condition
where the absorbed state of the polymer is {\em locally unstable}. Thus it identifies an upper bound
of instability, which is instructive to study. However, the actual desorption may take place at a smaller force. 
In the large $q$ limit, we have ${\cal A}_{11} \approx -[\k q^4+ \t q^2] $.
The two eigenvalues of matrix  ${\cal A}$ are
$
\l_\pm = \hf [{\cal A}_{11}+{\cal A}_{22} \pm \sqrt{({\cal A}_{11}+{\cal A}_{22})^2 -4 ({\cal A}_{11} {\cal A}_{22} - {\cal A}_{12} {\cal A}_{21})} ].
$
The condition that $\l_+ > 0$ for the mode to be unstable is satisfied
if $ {\cal A}_{12} {\cal A}_{21} - {\cal A}_{11} {\cal A}_{22} > 0$.

To obtain a closed analytic form for the expression of instability condition, we linearize the force dependence of detachment rate $\off \approx \w_0 (1+F_z/N_b f_d)$, where $N_b = n_b L$. 
This assumption is reasonable in the weakly bending limit where a large number of motor proteins would be in the bound state. 
Note that in the absence of external force, bound motor density $n_b^0 = \r \o_d$, with the duty ratio of motor proteins,  $\o_d = \on/[\on +\w_0]$.
Assuming $\s$ as the smallest length scale in the problem, so that $q \sim 1/\s$, the condition $\l_+ > 0$ leads to an inequality identifying a force which will destabilize the steady state profile 
of an adsorbed filament. 
Thus the following expression gives the critical desorption force,
\bea
F_z^c \approx \f{b^{\prime}_0f_dN_b}{\sqrt{1-\o_d}} \left[ 1 - b_1^{\prime} \f{\o_dk_m\, n_u \s^4}{\k + b^{\prime}_2v_0 + b^{\prime}_3 }  \right]^{1/2},
\label{Fztot}
\eea
using $|v_t| \approx v_0$ in the limit of small $F_z/N_b \ll f_s$. 
Here $b^{\prime}_1 = \s b_1$, $b^{\prime}_2 = \s^2b_2$ and $b_3^{\prime} = \s^4b_3$.  Note that the actual desorption may occur at force values smaller than the destabilizing force obtained from linear stability analysis, and thus we introduce the  proportionality constant $b^{\prime}_0$ in the above expression in order to compare it with numerical simulations.
 

The above expression shows how the critical desorption force $F_z^c$ is expected to depend on various properties of the system, 
like duty ratio $\o_d$, bending stiffness $\k$, and motor velocity $v_0$. 
The simulation data for adsorption-desorption phase diagrams fits well with Eq.(\ref{Fztot}) [see Figs (\ref{fzfs}) and (\ref{fzkappa})\,]. In the limit of $v_0=0$ and $\o_d$ following an equilibrium 
on-off process due to a sticky surface, the above expression describes passive desorption of a stiff filament~\cite{Kierfeld2006}.
%
In the limit of large $v_0$ and $\k$, the relation simplifies to 
\bea
F_z^c \approx b_0^{\prime}f_d N_b / \sqrt{1-\o_d}
\label{FzOd}
\eea 
where $N_b$ denotes the total number of bound motors at the onset of instability.
This expression contains only one unknown parameter $b_0'$, and thus
we shall present fitting of this expression with simulated data obtained in the relevant limit. As it turns out, the simulated phase diagram, 
in the large $v_0$ and $\k$ limit, is captured well by the expression obtained above [Fig.~\ref{fzfs}(c)].


\section{Outlook} 
Using mean field theory and linear stability analysis in one hand, and a stochastic MD simulation in the other, we investigated the adsorption-desorption transition of a semiflexible polymer 
attached to and actively  driven by a bed of molecular motors.  We have shown that the non-equilibrium transition is arguably a continuous transition. This is characterized by a gradual change in the fraction of bound motors or desorbed length with increasing pulling force, an absence of phase coexistence, and increasing correlation time as one approaches the critical point. 
We obtained the dependence of desorption force $F_z^c$ as a function of the bending rigidity $\k$, duty ratio $\o_d$  and active velocity $v_0$.  Phase diagrams obtained from detailed numerical simulations showed good agreement with our theory. 

The model we studied is closely related to microtubule (MT) organization in animal cells, particularly those MTs which grow from the microtubule originating centers (MTOC) towards the 
cell membrane, and get captured by the membrane associated dyenein motors. These motors grab the MT, and tries to walk towards the MTOC by pulling MTs towards the cell membrane. Qualitative 
understanding from our study still remains valid in such scenarios. 

Further, our model may be extended to understand cell adhesion in presence of elastic relaxation of cell membranes, as opposed to the rigid membranes considered in the seminal work by Bell~\cite{Bell1978}.
This might be achieved by considering two semiflexible filaments, as one dimensional projection of two dimensional membranes, and replacing the irreversibly attached motor proteins by freely diffusing reversible bonds.

Our choice of biologically relevant parameter values makes the current study an interesting prospect for experimental verification, e.g., in microtubule-kinesin gliding assays. Variation of $v_0$ and $\o_d$ may be achieved by changing ATP concentration. Bending rigidity $\k$ is partially tunable changing the ambient electrolyte concentration. Work is on to extend our model to study two-dimensional collective motion of semiflexible filaments driven by molecular motors. Particular questions as to how defects in activity of molecular motors~\cite{Scharrel2014} impact motility of single polymers and in turn the collective motion, will be studied. 

\acknowledgments
This work was initiated from a discussion of DC with  Frank J{\"u}licher and Debashish Chowdhury of IIT-Kanpur.  We thank Frank J{\"u}licher for valuable discussions, and
Guillaume Salbreux for  useful comments and a  critical reading of the manuscript.

\appendix

\begin{figure}[t]
\begin{center}
\includegraphics[width=0.9\linewidth]{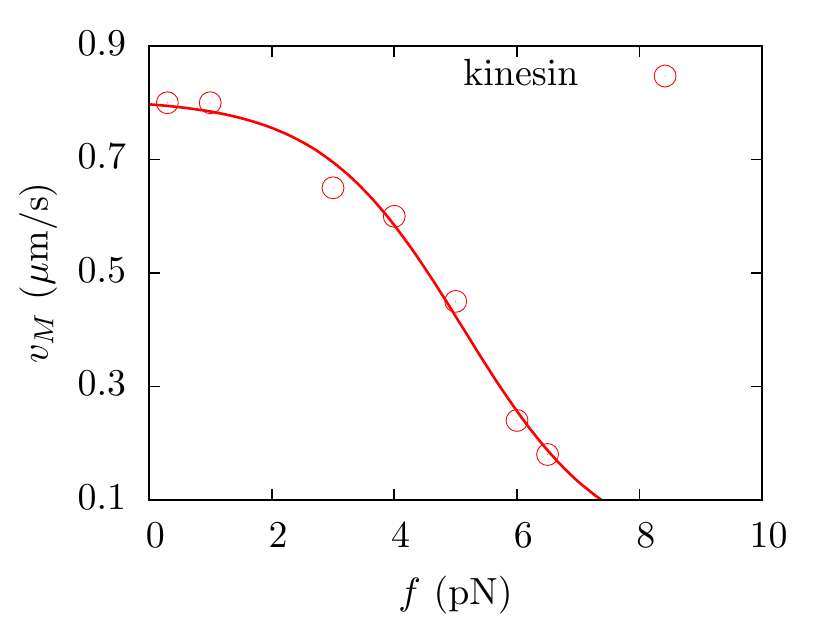} 
\caption{(Color online) Force-velocity data for kinesin molecules at 2 mM ATP concentration extracted from Ref.~\cite{Schnitzer2000a}. The line is a fit to
Eq.~\ref{eq:vm} with $v_0=0.807 \,\mu{\rm m\, s}^{-1}$, $d_0 = 0.01$ and $f_s = 1.16$ pN.  
}
\label{fig:cartoon}
\end{center}
\end{figure}

\section{Non-linear force velocity relation}
\label{appendixB}

The dependence of the velocity of procesive kinesin motor on the ambient ATP concentration
was successfully reproduced by Michaelis-Menten kinetics~\cite{Schnitzer2000a}. This describes the binding of
kinesin motor head M (enzyme) to an ATP molcule (fuel) and the subsequent ATP-hydrolysis

\begin{displaymath}
{\rm M + ATP} \autorightleftharpoons{$r_a$}{$r_d$} {\rm M.ATP} \autorightarrow{$r_h$}{} {\rm M + ADP + P_i}.
\end{displaymath}
Here $r_a$ ($r_d$) is rate constant for binding (unbinding) of an ATP to the kinesin head, and
$r_h$ is the rate constant of ATP-hydrolysis; ${\rm P_i}$ denotes the phosphate ion.

This leads to the Michaelis-Menten expression for the motor-velocity,
\beq
v(f) = d\, r_h(f) \Psi([ATP]) 
\eeq
where $d$ is the step-size by which the molecular motor moves per ATP hydrolysis, 
$K_M=(r_h+r_d)/r_a$ is the Michaelis-Menten constant, $\Psi ([ATP]) = [ATP]/([ATP]+K_M)$ with
$[ATP]$ denoting the ATP concentration. 
%

The net time scale of ATP hydrolysis has two components, one is the time required in absence of force $t_1$,
the other one is an exponentially increased time scale $t_2 \exp(f\d/\kb T)$ to
cross the enhanced energy barrier $f\d$ ($\d$ is a characteristic molecular length scale)
using thermal energy $\kb T$. Thus
the total time per ATP hydrolysis is $t(f)=t_1+t_2 \exp(f\d/\kb T)$
with the corresponding rate $r_h(f)=1/t(f)$. This leads to
the following general form of the rate of ATP hydrolysis \cite{Schnitzer2000a},
\bea
r_h(f) = \f{r_h(0)}{1+d_0 \exp(f/f_s)} 
\eea
where $f_s=\kb T/\d$, $r_h(0)=1/t_1$, $d_0=t_2/t_1$.
The load-free sliding velocity of a single motor is coupled to ATP-hydrolysis by
$v_0=r_h(0)d\,\Psi([ATP])$. Self propulsion $v_0$ is thus a function of ATP concentration, and at 
high enough concentration saturates to $v_0=r_h(0)d$.
Thus  the motor velocity 
\beq
v_M(f)=\f{v_0}{1+d_0 \exp(f/f_s)}.
\label{eq:vm}
\eeq
Fitting this form with kinesin force-velocity data obtained at large ATP concentration of 2 mM gives $v_0=0.807 \,\mu{\rm m\, s}^{-1}$, 
$d_0 = 0.01$ and $f_s = 1.16\,$pN~\cite{Schnitzer2000a}.  The maximal force generated by single kinesin molecule is $\a v_M(f)$ where $\a$
denotes the viscosity of ambient medium. We used this expression and above-mentioned parameter values 
to model active force generation of molecular motors in the main text.  


Note that the above-mentioned chemical reaction describes 
the motion of motor head when it is attached to the polymer track. 
In the detached state, it performs simple diffusion.

\begin{figure}[t] 
\begin{center}
\includegraphics[width=0.9\linewidth]{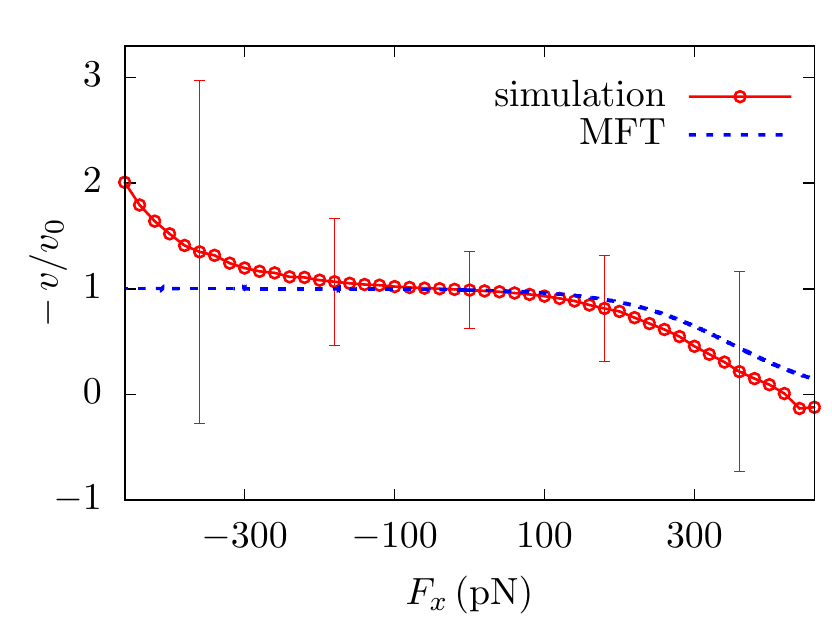} 
\caption{(Color online)
Velocity of the polymer as a function of external force in one-dimension. 
Error bars indicate numerical errors at some representative values of simulation data. The dashed line is the MFT prediction.
}
\label{vxfx}
\end{center}
\end{figure}

\section{1D characterization}
\label{appendixA}

We perform stochastic MD simulations, of an one dimensional (1D) rigid rod, under external force $F_x$, and activity of the molecular motors. When attached
to the filament, the head of molecular motors walk towards positive $x$-direction with active velocity $v^a$ while the tail remains irreversibly attached to the substrate, 
thereby pushing the filament towards the opposite direction. A positive $F_x$ thus acts like an
opposing load to the motor driven motion of the filament, whereas a negative $F_x$ assists that motion. The stochastic noise acts on the filament as a whole to maintain the ambient 
temperature. In the following, we analyze the motion using mean field theory (MFT) in the over damped limit, and compare the predictions with simulation results.

The {\em head} of each motor protein, when
attached to the filament may move dragged by the filament moving with velocity $v$, or due to its active relative motion $v^a$. Thus with respect to the attachment point on the substrate, the
extension of the {\em head} position is 
\bea
\f{d x}{ d t} = v + v^a.
\eea
The resultant force on the filament due to $N_b$ attached motors is 
\bea
f^a = - k_m x\, N_b. 
\label{eq:fa}
\eea
The filament velocity is given by the force balance
\bea
\a v = f^a + F_x.
\label{eq:v}
\eea
The number of bound motors obey the master equation 
\bea
\p_t N_b = \on N - (\on +\off ) N_b,
\eea
 where $N$ is the total number of motor proteins. At steady state, 
$\p_t N_b=0$ implies 
\bea
N_b = N\, \on/(\on + \off),
\label{eq:nb}
\eea
and $d x/d t=0$ implies a negative velocity of the filament $v = -v^a$. On an average, the filament moves in the direction opposite to the motors with the velocity of a single free
motor protein. 

\begin{figure}[t] 
\begin{center}
\includegraphics[width=0.9\linewidth]{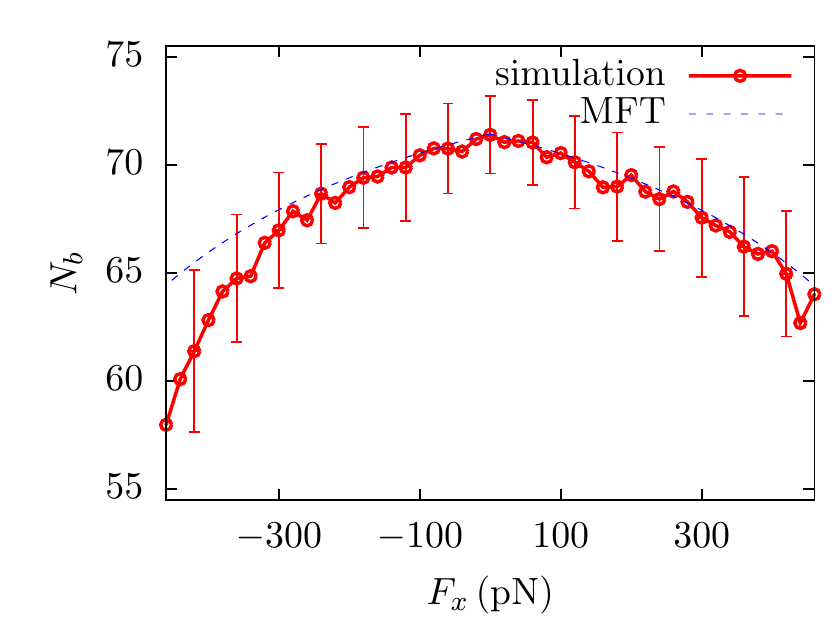} 
\caption{(Color online)
Average number of attached motors as a function of external force in one-dimension. We show error bars at representative numerical data. The dashed line shows the 
MFT prediction.
}
\label{nbfx}
\end{center}
\end{figure}

In the presence of external force, 
the motor detachment rate 
$\off (x) = \w_0 \exp(k_m |x|/f_d)$, and the active velocity $v^a (x) = v_0 / [1+d_0\exp(k_m x/f_s)]$. 
Using Eq.\ref{eq:fa} and  $v=-v^a$, Eq.\ref{eq:v} leads to   
\bea
k_m x = \f{F_x + \a v^a(x)}{N_b(x)},
\label{eq:ky}
\eea
where $N_b$ can be expressed as a function of $x$ via Eq.\ref{eq:nb} and $\off (x)$. The non-linear algebraic relation Eq.\ref{eq:ky} can be solved for $x$. This in turn
gives the value of $v^a(x)$ and therefore the filament velocity $v$ at a given value of $F_x$. 
This further allows us to calculate $N_b$ as a function of $F_x$ through Eq.~\ref{eq:nb}. 

Using the parameters $\on=20\,{\rm s}^{-1}$ $\w_0 = 1{\rm s}^{-1}$, 
$\a=1\,{\rm pN\,s\,}\mu{\rm m}^{-1}$, $d_0=0.01$, $f_s=1.16\,$pN, $f_d=6\,$pN, 
and setting the total number of available motor proteins $N=75$
we plot the $F_x$ dependence of $v$ and $N_b$ in Figs.~\ref{vxfx}, and ~\ref{nbfx} respectively. 
The plots show comparison of this MFT prediction with simulation results, and we find reasonably good agreement over a broad range of $F_x$. 
The number of attached motors $N_b$ reduces with increase in the load force, and thus is independent of the sign of $F_x$. 
In absence of external force, the filament moves with velocity $v=-v_0$ as expected, and
the speed reducing with increasing opposing load $F_x>0$. However for assisting load $F_x<0$, MFT predicts a $v$ independent of $F_x$. 
Though for $F_x>0$ we see good agreement between MFT and simulations, for large negative force $F_x$ we find qualitative
deviation. At large external forces,  the steady state assumption $dx/dt=0$ does not hold, and velocity of the filament is expected to be $v \sim F_x$ independent of the 
active force. The deviation from MFT shows a precursor of this crossover.

\bibliographystyle{prsty}
\bibliography{wlc_motor}
\end{document}